\documentclass{ws_jk}

\def\NPBP{{\em Nucl. Phys.} B (Proc. Sup.)}

\begin{document}

\title{From Surface Roughening to QCD String Theory}

\author{K. Jimmy Juge}

\address{Fermi National Accelerator Laboratory, 
 P.O.Box 500, Batavia, IL 60510, USA\\
 E-mail: juge@fnal.gov}

\author{Julius Kuti\thanks{ Speaker at the 24th Johns Hopkins Workshop
on Nonperturbative QFT and Their Applications, Budapest, Hungary, August 19-21,
2000.}}

\address{University of California at San Diego,
La Jolla, CA 92093, USA\\ 
E-mail: jkuti@ucsd.edu}

\author{Colin Morningstar}

\address{Carnegie Mellon University,
Pittsburgh, PA 15213, USA\\
E-mail: cmorning@andrew.cmu.edu}

\maketitle

\abstracts{Surface critical phenomena and the related onset of 
Goldstone modes represent fundamental properties of the confining flux
in Quantum Chromodynamics. New ideas on surface roughening and their
implications for lattice studies of quark confinement and string formation 
are presented. Problems with a simple string description of the 
large Wilson surface are discussed.}

\section{Introduction}

There exists great interest and considerable effort to explain quark 
confinement in Quantum Chromodynamics (QCD) 
from the string theory viewpoint. The ideas of
't Hooft, Polyakov, Witten, and others, and 
recent glueball spectrum or QCD string tension
calculations in AdS theories are some illustrative examples of these activities.
In a somewhat complementary approach, the search for a microscopic
mechanism to explain quark confinement in the QCD vacuum continues with
vigorous effort. 
It is useful to note (in loose chronological order) some of the ideas
on QCD string formation:
\vskip 0.1in

\begin{description}

\item[(i)] {\em The strong coupling lattice picture.} 
This is the oldest ``string-like"
confinement picture which was formulated in the early days of lattice
QCD and immediately raised the issue of the surface roughening 
transition on a large area Wilson loop.

\item[(ii)] {\em Microscopic confinement mechanisms.}
To understand confinement as we move past the roughening
transition from strong coupling towards the
continuum limit, on-lattice (and off-lattice) ideas were developed
about microscopic confinement mechanisms in the QCD vacuum. 
Candidates of dominant gauge field configurations
include instantons, monopoles, Z(N) flux configurations, and other
examples.
A popular idea is based on dual superconductivity with magnetic monopoles
playing the dominant role in an underlying effective
Landau-Ginzburg type low energy theory of the confining flux.

\item[(iii)] {\em Large N expansion.} 't Hooft suggested that in
the large N limit of nonabelian SU(N) gauge theories, the summation
of the leading diagrams might lead to the expected 
QCD string picture of quark confinement. 
This idea is very powerful and remains much studied
today in various settings.

\item[(iv)] {\em QCD in loop space formulation}. 
Polyakov 
reformulated the path integral approach to nonabelian Yang-Mills fields
with the hope that ``string-like" microscopic variables will help
to understand the emergence of the confining string.

\item[(v)] {\em Confinement from higher dimensions.} 
This very popular idea is based on
the higher dimensional Anti-deSitter (AdS) space and one of its ambitious
goals is to the understand the origin of quark confinement.

\item[(vi)] {\em Surface spectrum and D=2 conformal field theory.} 
We will argue 
that a deeper understanding of the string theory connection
with large Wilson surfaces will require a more precise knowledge
of the surface excitation spectrum and the determination of 
the appropriate universality class of surface criticality. This new
QCD string universality class will also require a consistent
description of the conformal properties of the gapless surface 
excitation spectrum.

\end{description}

\section{Lattice Study of QCD String Formation: Three Objectives}

In this progress report we present our ab initio on-lattice 
calculations to probe the dynamics
of string formation in QCD when the separation of the static 
quark-antiquark pair at the two ends of the confining flux becomes large.
Our view is that the relevant properties of the underlying 
effective QCD string theory,  
whether it emerges from strictly string theoretic ideas, or from the microscopic
theory of the confining vacuum, are coded in the excitation spectrum of the
confining flux. To establish the main features of this spectrum remains 
our main objective. 
In collaboration with Mike Peardon\cite{MP}, we are also developing the relevant
methodology to study the spectrum of a ``closed" flux loop 
across periodic slab geometry (Polyakov line) by choosing
appropriate boundary conditions and operators for selected excitations
of the flux without static sources. 

Throughout this work we will focus on the confining
properties of the nonabelian Yang-Mills field and the effects of light
quark vacuum polarization will be neglected. 
Since this is an interim progress report,
we will focus only on the
results of the calculations and their physical interpretation.
The technical details will be completely omitted.
Acting within space and time limitations,
we will be unable to provide appropriate references
to the literature. These omissions will be remedied in our forthcoming
publications. 
As reported here in the next
three sections, the main thrust of our recent work
develops along three closely related lines of investigations:

\begin{description}

\item[~~~~]{\bf Section 3:~}{\em The Excitation Spectrum of the Wilson
Surface in QCD.}
\newline After building a Beowolf class UP2000 Alpha cluster with 9.33 Gflops 
computing power,\cite{JK} dedicated to this project, 
we increased the statistics of our earlier
work on the excitation spectrum of the Wilson surface by more than an order
of magnitude. We determined the spectrum as the
function of the ${\rm Q\overline Q}$ separation and established three well
identified scales of the confining flux. We also found that
the main features of string formation share some universal
properties, independent of the tested gauge groups SU(2) and SU(3),
and the tested space-time dimensions D=3 and D=4.

\item[~~~~]{\bf Section 4:~}{\em Surface Excitation Spectrum in 
Z(2) Spin and Gauge Models.}
\newline A quantitative analysis of the exact surface excitation
spectrum and its conformal properties are  presented for the BCSOS model 
in three dimensions.
From the Bethe Ansatz equations we calculate numerically 
the exact spectrum of the interface using transfer matrix methods. 
We also interpret
this spectrum in two-dimensional conformal field theory. 
The surface physics of the Z(2) gauge model is closely related to the BCSOS
model by universality argument and a
duality transformation. The string limit of the confining flux
in the Z(2) model will be discussed in the critical region of the bulk.
New issues will be raised about the crossover behavior of the confining
Wilson surface in the Z(2) gauge model as we move from the roughening
transition into the critical domain of the bulk embedding medium of the
surface. Some puzzling features remain unresolved as we progress.

\item[~~~~]{\bf Section 5:~}{\em What Is the Continuum Limit
of QCD String Theory?}
\newline The intricate connection between 
the on-lattice roughening
transition of the Wilson surface and the crossover to continuum 
string behavior will be discussed in ${\rm QCD_3}$.
Based on the investigation of the Z(2) gauge model,
we describe the important crossover
issue as we move from the roughening transition point of 
the Wilson surface towards the critical domain of the gauge theory
vacuum which represents the bulk embedding medium for the Wilson surface
in the continuum limit.
We will raise the question whether the universality class of the
Wilson surface in the continuum is different from the Kosterlitz-Thouless
universality class of the surface at the roughening transition.

\end{description}

\section{The Excitation Spectrum of the Wilson Surface in QCD}

A rather comprehensive determination of the rich energy spectrum
of the gluon excitations
\begin {figure}[h]
\vskip -0.2in
\epsfxsize=4.5in
\epsfbox{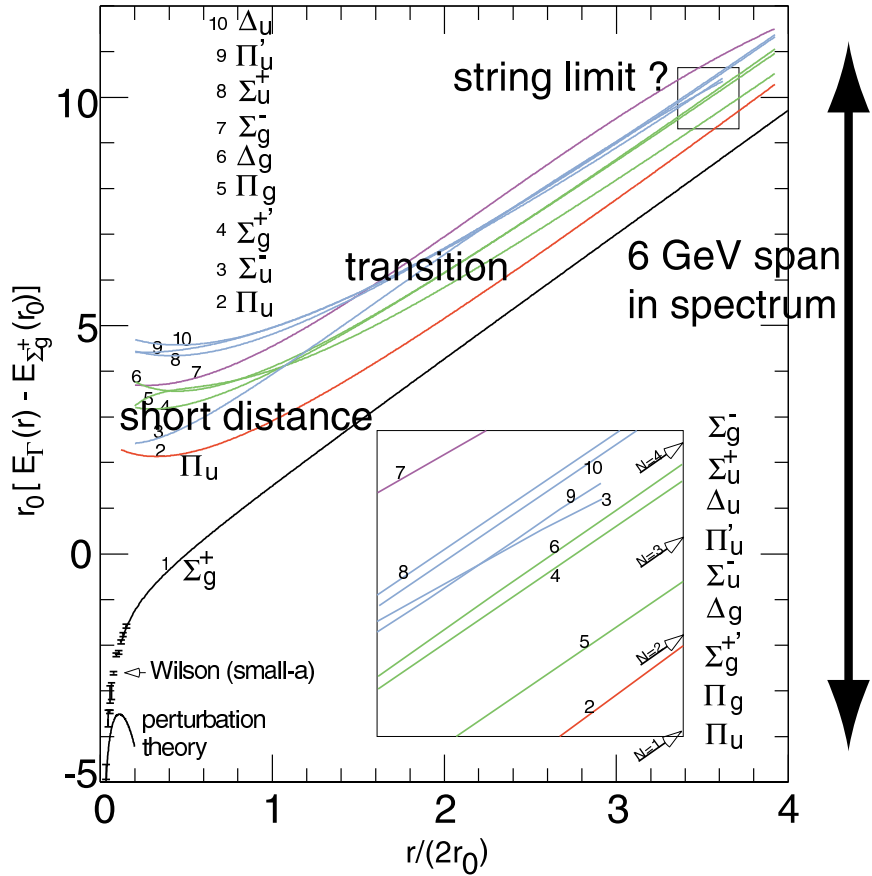}
\caption{Continuum limit extrapolations are shown for
the excitation energies where an arbitrary constant is removed
by subtraction. 
Color coding in postscript is added to the numerical labelling of
the excitations (N=0,black:1), (N=1,red:2), (N=2,green:4,5,6), 
(N=3,blue:3,8,9,10),
and (N=4,cyan:7). The five groups represent the expected quantum numbers
of a string in its ground state (N=0) and the first four excited states
(N=1,2,3,4).
The arrows in the inset represent the expected locations of the four lowest 
massless string excitations (N=1,2,3,4) which have to
be compared with the energy levels of our computer simulations.}
\label{fig:fig1}
\vskip -0.2in
\end{figure}
between static sources in the fundamental representation of ${\rm
SU(3)_c}$ in D=4 dimensions was reported earlier\cite{JKM1,JKM2}
for quark-antiquark separations r ranging from 0.1 fm to 4 fm. 

The extrapolation
of the full spectrum to the continuum limit
is summarized in Fig.~\ref{fig:fig1} with very different
characteristic behavior on three separate physical scales.
Error bars are not shown, and the earlier results displayed here remain
compatible with our new run on the UP2000 Alpha cluster after more than
a tenfold 
increase in statistics (a Bayesian statistical analysis on the new
results is in progress). 
Our notation and the origin of the quantum numbers used in 
the classification of the energy levels are explained in the 
Appendix.
Following Sommer\cite{Sommer},
the physical scale ${\rm r_0}$	in Fig.~\ref{fig:fig1} is
set by the relation
\begin{equation}
{\rm [r^2 dE_{\Sigma^+_{\rm g}}(\vec{r})/dr]}_{\rm r=r_0}=1.65~.
\label{eq:sommer}
\end{equation}
This scale turns out to be ${\rm r_0=0.5~ fm}$ to a good approximation.

The full spectrum, determined as a function of quark-antiquark separation,
spans over 6 GeV in energy range which requires rather sophisticated
lattice technology.
Qualitatively, we can identify three distinct regions in the spectrum.
Nontrivial short distance physics
dominates for $r \leq 0.3~{\rm fermi}$. The transition region 
towards string formation is identified on the scale $0.5~{\rm fm}\leq
2.0~{\rm fm}$. String formation and the onset of
string-like ordering of the excitation energies
occurs in the range between 2 fm and 4 fm which
is the current limit of our technology.

\subsection{Short Distance Physics}
 
The observed energy levels, even their qualitative ordering,
are in violent disagreement with naive
expectations from a fluctuating string for quark-antiquark separations 
${\rm r\leq 2~ fm}$. It is not difficult to show that for small 
${\rm r\leq 0.3~\rm fm}$ the
non-string level ordering is consistent with the short distance operator 
product expansion (OPE) around static color sources for gluon excitations.
Consider the static QCD Hamiltonian in Coulomb gauge,

\begin{equation}
{\rm 
H_{Cb} = H_{gluon} + \frac{g^2 Q\cdot 
\overline{Q}}{4\pi  \mid \vec{r}_1 - \vec{r}_2 \mid} 
- g^3(Q\times \overline{Q})_a\int d^3r\vec{A}_a(\vec{r},t)\vec{J}(\vec{r})
}
\label{eq:coulomb}
\end{equation}
where the intrinsic color operator ${ \rm Q\cdot\overline{Q} }$ takes
the value -4/3 for a color singlet pair, and 1/6 for the color octet 
state of the sources;
${\rm \vec{A}_a(\vec{r},t) }$ is the transverse gluon field operator,
and a summation is understood over the color index a=1,2,...,8.
The transverse current, 

\begin{equation}
{\rm
4\pi\vec{J}(\vec r) = ( \frac{1}{\mid \vec{r} - \vec{r}_1 \mid}
- \frac{1}{\mid \vec{r}_1 - \vec{r}_2 \mid} )
\widehat{\nabla}
( \frac{1}{\mid \vec{r} - \vec{r}_2 \mid}\
- \frac{1}{\mid \vec{r}_1 - \vec{r}_2 \mid} ),
}
\label{eq:current}
\end{equation}
where the symmetric $\widehat{\nabla}$ operator 
acts on both sides on ${\rm \vec{r}}$,
couples the static field of the color sources (localized
at positions ${\rm \vec{r}_1}$
and ${\rm\vec{r}_2}$) to the transverse gluon field 
${\rm \vec{A}_a(\vec{r},t)}$.
The Hamilton operator of Eq.~(\ref{eq:coulomb}) is the starting point
of the OPE and the somewhat more phenomenological bag model.\cite{JKM3,soto}
They both imply that it is sufficient to keep the leading terms
of the multipole expansion in Eqs.~(\ref{eq:coulomb},\ref{eq:current})
at short distances.
The OPE is quite general and it is expected to break down around 
${\rm r \approx 0.2-0.3~fm}$ whereas the bag model can be
extended to larger r values by simply keeping all multipoles of a more
specific {\em confined} and static chromoelectric field
in the numerical solution of the coupled equations of the
static sources and the transverse gluon field.\cite{JKM3,JKM4}

One of the remarkable predictions of the short distance physics
is the expectation that several groups of gluon excitations
\begin {figure}[h]
\vskip -0.3in
\epsfxsize=4.5in
\epsfbox{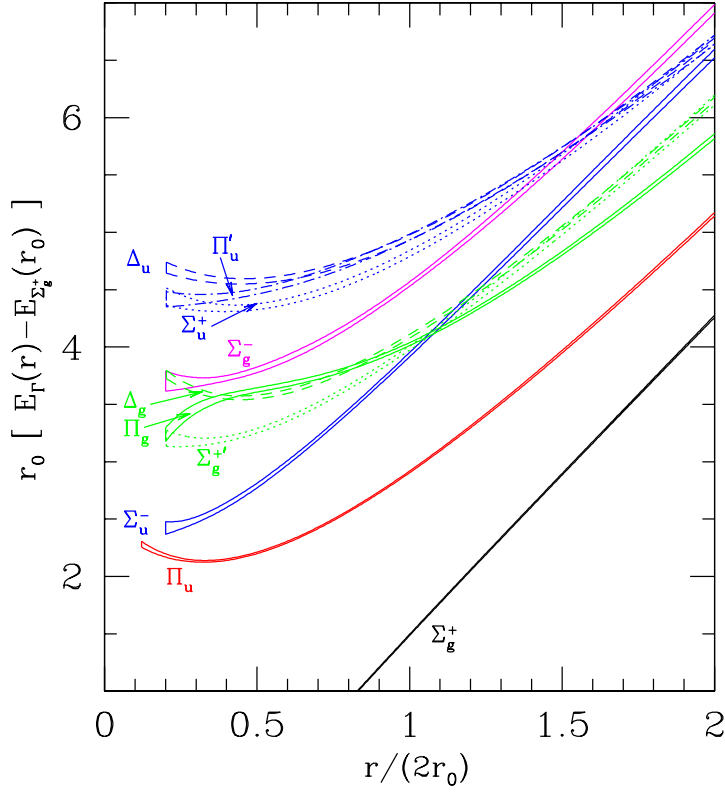}
\vskip -0.1in
\caption{Continuum limit extrapolations with error range are shown 
for three groups of nine excited states to illustrate the prediction
of the short distance operator product expansion.}
\label{fig:fig2}
\vskip -0.2in
\end{figure}
should be approximately degenerate. Fig.~\ref{fig:fig2} shows
the excitation spectrum on a magnified scale, and it includes some
of the additional states not displayed in Fig.~\ref{fig:fig1} 
to illustrate the approximate working of the
predicted degeneracies.
Continuum limit extrapolations with error range are shown 
for three groups of nine excited states: 
($ \Sigma^+_u,\Pi^{'}_u, \Delta_u $),
($ \Sigma^-_g,\Delta_g,\Sigma^{+'}_g,\Pi_g)$, and 
($ \Sigma^-_u,\Pi_u$).
Primes designate
``radial" excitations in the appropriate quantum number combinations.
The groups are predicted
to be approximately degenerate in the multipole expansion.\cite{JKM3,soto} 
This is based on the plausible physical picture that the confined
static field of the ${\rm Q\overline{Q}}$ pair is mostly dipole for
${\rm r \leq 0.3~fm}$ with the lowest terms dominating in the
operator product expansion.

Within each group one has different string quantum numbers mixed
together. For example,
$ \Sigma^-_u$ is N=3 and $ \Pi_u $ is N=1, 
therefore the two states are expected 
to split into separate string levels for large r. Also, the 
$ \Delta_g,\Sigma^{+'}_g,\Pi_g$ states should be degenerate
and sandwiched between the
$ \Pi_u $ state and the
$ \Sigma^-_u$ state in the string limit, with equidistant spacing in-between.
Instead, they are located higher than the $ \Sigma^-_u$ 
state at short distances, and become 
degenerate with the $ \Sigma^-_g $ state which is an N=4 string
excitation! On the intermediate scale for ${\rm 0.5 ~fm\leq r \leq 2~fm}$,
a remarkably rapid rearrangement of the energy levels is observed.

We should  also note that for r above 0.5 fm,
all of the excitations shown in Fig.~\ref{fig:fig1}
are stable with respect to glueball decay.
As r decreases below 0.5 fm, some of the excited levels eventually become
unstable resonances as their energy gaps above the ground state 
$\Sigma_g^+$ exceed the mass of the lightest glueball.
They require more care in their interpretation.

\subsection{Transition Region for $ 0.5~fm\leq r \leq 2~fm$}

One of the striking features of the transition region is 
the dramatic linearly-rising behavior of the $\Sigma_{\rm g}^+$ 
ground state energy in Fig.~\ref{fig:fig1} 
once ${\rm r}$ exceeds about ${\rm 0.5~fm}$.
The empirical function
${\rm E}_{\Sigma_{\rm g}^+}({\rm r}) = -{\rm c}/{\rm r} + \sigma{\rm r}$
approximates the ground state energy very well for
${\rm r}\geq {\rm 0.1~fm}$ with the fitted constant ${\rm c = 0.3}$.
Early indoctrination on
the popular string interpretation of the confined flux
for ${\rm r}\leq{\rm 1~fm}$ was mostly based on the observed shape of the
$\Sigma_{\rm g}^+$ ground state energy and some rudimentary
determination of a few excited states.\cite{michael} The
linear shape of the ground state
potential for ${\rm r} \geq~{\rm 0.5~fm}$ and the approximate agreement
of the curvature shape for ${\rm r} \leq 0.5~{\rm fm}$ with the ground
state Casimir energy $-\pi/(12{\rm r})$ of a long
confined flux\cite{Luescher} was interpreted as evidence for 
string formation.
Our observed excitation	spectrum clearly contradicts claims on
the simple string interpretation of the linearly rising confining
potential for ${\rm r\leq 2~fm}$.
The observed energy levels, even their qualitative ordering,
are in gross disagreement with
expectations from a fluctuating string for quark-antiquark separations 
${\rm r\leq 2~ fm}$. 

There is no solid theory for this transition region
which is perhaps the most difficult to describe. 
It is interesting to note, however, that the somewhat
phenomenological bag model does quite well in 
this transition region,\cite{JKM3}
which remains the most model dependent scale with
unknown microscopic details of the confinement mechanism
in the QCD vacuum.

\subsection{String Limit Between $ 2~fm\leq r \leq 4~fm$?}

Although the transition region is not string-like, 
the rapid rearrangement of the energy levels 
to reach string-like
level ordering around ${\rm r \approx 2~fm}$ is remarkable. 
For example, the states $ \Sigma^-_u$ and
$ \Sigma^-_g $ break away from their respective short distance degeneracies to
approach approximate string level ordering for ${\rm r \geq 2~fm}$
separation.
In general, when we reach ${\rm r\approx 2~fm }$ quark-antiquark 
separation, after a
rapid movement of the energy levels, 
a new level ordering emerges which qualitatively begins to resemble
a naive string-like spectrum which is anticipated on quite general
grounds. However, this new level ordering exhibits a finite structure
whose origin is puzzling and important to understand.

A very robust feature of the effective low-energy description of a fluctuating
flux sheet, or interface, in continuum euclidean space is the 
expected presence of massless Goldstone excitations associated with the
spontaneously-broken transverse translational symmetry.
The emergence of a QCD string theory from other theoretical
considerations would also suggest a massless excitation spectrum.	 
These transverse
modes have energy separations above the ground state given by multiples of
${\rm \pi/r }$ for fixed ends.  After the rapid transition,
the level orderings and approximate degeneracies
of the gluon energies at large r match, without exception, those expected
of the Goldstone modes.	 However, the
precise separation $\pi/{\rm r}$  of the energy levels
is not observed in our spectrum. Some of the expected
degeneracies are also significantly broken. This is likely to come
from the finite size scaling theory of the underlying effective string theory
which is governed by the higher dimensional operators of the effective
string Lagrangian and their effects on the spectrum. In that sense
the fine structure of our spectrum contains
the information about the underlying effective string theory.
These remarks will be illustrated in model examples of sections 4 and 5.

\subsection{D=3 and D=4 SU(2) Results}

The approximate string-like ordering of the energy levels between
${\rm 2~fm\leq r \leq 4~fm}$ and yet the substantial deviations
from the expected locations of the massless excitations is tantalizing.
Although we are beginning to understand that 
the complex patterns should emerge
from the finite size scaling analysis of massless excitations
(model examples will be given in the next section), checks on the
methodology of the simulation results is important. This reason alone
would motivate the tests where the original D=4, SU(3) runs were
repeated with SU(2) color group in D=3 and D=4 dimensions.
The D=3 SU(2) simulations have further significance. The high
accuracy of the results allows more detailed comparisons with
theoretical ideas which are derived from detailed studies of
three-dimensional SOS models and  gauge models. The fluctuating
Wilson surface of the D=3 SU(2) simulations is expected to share
some common features with interfaces of more simple three-dimensions 
gauge and spin models.

We turn now to the salient features of our SU(2) test results.
\begin {figure}[h]
\epsfxsize=3.8in
\epsfbox{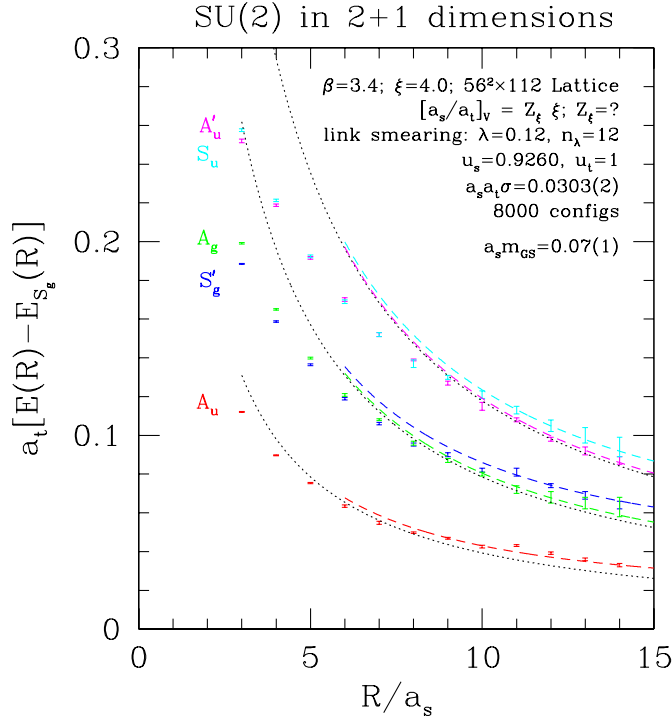}
\vskip -0.1in
\caption{The energy gaps above the ground state are shown for SU(2)
in D=3 dimensions. The symbols here refer to the point group representations
of the surface states.
The dotted lines are the expected locations of the 
massless Goldstone excitations. The dashed lines are ad hoc fits of the
data to massive string excitations which is further discussed in the text.}
\label{fig:fig3}
\vskip -0.2in
\end{figure}
The most striking feature of the D=3 and D=4 SU(2) simulations
(D=3 depicted in Fig.~\ref{fig:fig3}) is the universality of the
results:
\begin{description}
\item[(a)] 
The level orderings and approximate degeneracies
of the gluon energies at large r match, without exception, those expected
of the string modes for both gauge groups SU(2) and SU(3)
and for D=3, or D=4. 
\item[(b)]
Even the fine structure of the split spectrum
shows a great deal of universality in
the character of its substantial deviations 
from the expected massless spectrum at large
separations. First, in D=4 dimensions we 
observed that the deviations in the string formation region 
${\rm 2~fm\leq r \leq 4~fm}$ can be roughly described by 
replacing the expected massless excitation spectrum with an ad hoc fit to 
massive surface excitations parametrized by
a new scale parameter ${\rm m_s}$. 
Although the numerical value of ${\rm m_s}$ varies somewhat 
in the three cases we considered, this replacement alone describes 
qualitatively the splitting patterns 
which occur in all simulations for string quantum numbers which 
otherwise should be degenerate. The significance of the apparent
mass term is unclear. Very likely, it is only an artificial
description of finite size power corrections to the massless
string spectrum. More work is needed to clarify this, because a massive
QCD string is not an entirely excluded possibility. 
\item[(c)]
One more important test was done in our simulations.
We rotated out the static quark-antiquark pair from the main axis of the
lattice and determined the spectrum again in the diagonal off-axis position.
All the spectra we determined remained the same within error bars. 
This is significant for later discussions. It supports the
the plausible argument that we are sitting in the
rough phase of the large Wilson surface at the values of the coupling
constants where our extensive simulation results were obtained.
\end{description}

\subsection{Strong Coupling Tests}

\begin {figure}[h]
\epsfxsize=4.5in
\epsfbox{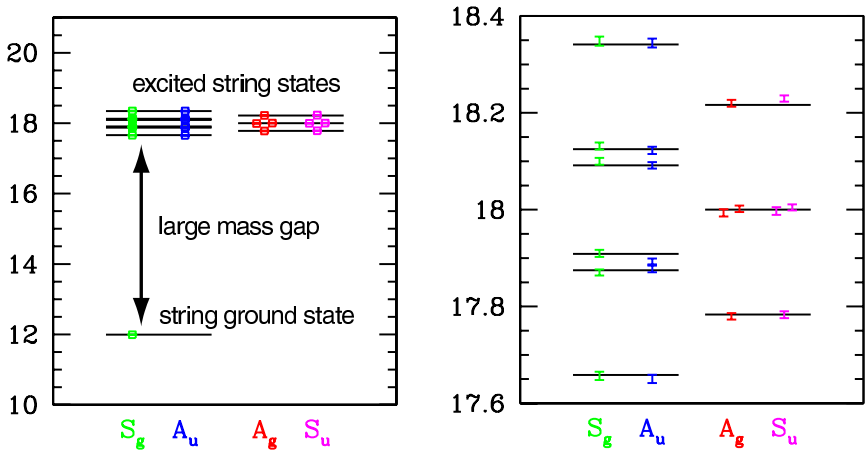}
\caption{Strong coupling tests of massive ``string"
excitations for D=3 in SU(2). The solid lines are calculated analytically
and the simulation results for 20 states sit within errors. The large
energies required huge lattice aspect ratios which is the most
challenging part of our lattice technology. The scale is set by the lattice
spacing a.}
\label{fig:fig4}
\end{figure}

To make sure that the rather sophisticated lattice technology we applied
is working we determined the excitation spectrum in extensive test runs 
at strong coupling in three-dimensional SU(2). 
This gave us the opportunity to compare
analytic results with our lattice technology. We calculated the string
spectrum in leading nontrivial order for r=3a and r=4a separations of the
static sources, on axis, at very strong coupling where the next to
leading correction is expected to be in the percent range. Fig.~\ref{fig:fig4}
displays the simulation results and the analytic predictions which are in
excellent agreement. This is a very nontrivial test supporting
our belief that the peculiar fine structure of our results in
the string formation region of ${\rm 2~fm\leq r \leq 4~fm}$ is
not an artifact of the simulation method which performs so well
under very difficult conditions.

\section{Surface Excitation Spectrum in Z(2) Spin and Gauge Models}

One of our extensive tests above included a detailed study of the
Wilson surface excitation spectrum of the D=3 SU(2) gauge model of
${\rm QCD_3 }$. The Abelian subgroup Z(2) of SU(2) is expected
to play an important role in the microscopic mechanisms of quark
confinement suggesting that
Wilson surface physics of the D=3 Z(2) gauge spin model should
have qualitative and quantitative 
similarities with the theoretically more difficult 
${\rm QCD_3 }$ case. 
In the critical region of the Z(2) model we have
a rather reasonable description of continuum string formation
based on the excitation spectrum of a semiclassical defect line 
(soliton) of the equivalent ${\rm \Phi^4}$ field theory. This
is the analogue of the effective Landau-Ginzburg equations of QCD.
It is useful to investigate surface physics in the BCSOS model, first.
The surface physics of the Z(2) gauge model is closely related to the BCSOS
model by universality argument and a duality transformation.

\subsection{Interface Spectrum in the BCSOS Model}
\begin {figure}[h]
\epsfxsize=2.5in
\epsfbox{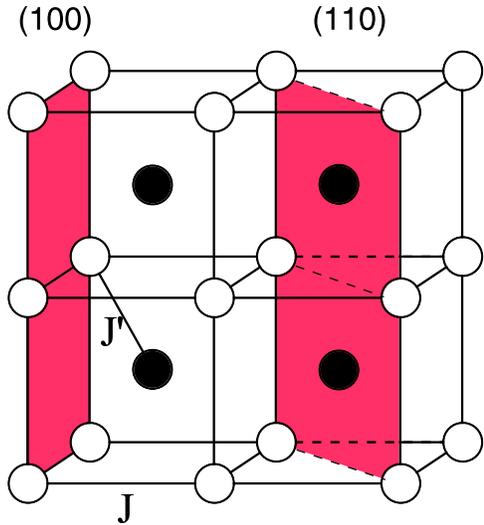}
\caption{The two couplings J and J' of the BC Ising lattice
are defined in the text.}
\label{fig:fig5}
\end{figure}

This model was proposed by Beijeren\cite{BJ} as the simplest interface
which can be investigated by analytic methods. We start from a body-centered
cubic Ising model with ferromagnetic nearest neighbor coupling J' (between
sites at the center and on the corner of an elementary cube as illustrated
in Fig.~\ref{fig:fig5}). The next-nearest-neighbor coupling J is defined
along the three main lattice directions. An interface can be created 
in the horizontal direction by keeping the spins in the two bottom layers
positive and those in the two top layers negative. Free, or periodic boundary
conditions can be imposed on the side walls of the 
finite three-dimensional lattice. 
The body-centered solid-on-solid (BCSOS)
model is obtained by letting J' approach infinity and keeping J constant.
The SOS condition is satisfied in this limit and the fluctuating interface,
with a height measured from a horizontal base,
separates the negative spin phase from the positive one.
For roughening studies, the BCSOS limit is expected to differ from the 
symmetric J=J' BC Ising model only by a few percent, since only very few wrong 
phase bubbles exist in the bulk around the roughening transition of the
fluctuating interface. 

The BCSOS model can be mapped into the six-vertex model for which
the Bethe Ansatz equations are known.\cite{EL} Beijeren showed from
the Bethe Ansatz that the BCSOS model has a phase transition at 
${\rm T_R = J/(k_B\cdot 2ln2)}$ which is identified as 
the critical temperature of the roughening transition
of the interface. For ${\rm T < T_R}$ the interface is smooth
with a finite mass gap in its excitation spectrum. For ${\rm T \geq T_R}$ 
the mass gap vanishes and the interface exhibits a massless
excitation spectrum. The roughening transition is of the
Kosterlitz-Thouless type in the model.

The connection with the Z(2) gauge model is rather straightforward.
The simple cubic Ising model is in the same universality class as
the body-centered Ising model. It follows then that the SOS limit
of the simple cubic Ising model should be very similar to
the BCSOS model. In fact, we expect that the low energy spectra
of the two interfaces should exactly map into each other, after
appropriate rescaling of the temperatures. In addition, we expect 
the SOS limits of both models to differ only
by a few percent from the interfaces of the original models.
Therefore the BCSOS interface should behave essentially the same as the
interface of the simple cubic Ising model.
As the final step of the
transformation, we note that the simple cubic Ising model can be mapped
by a duality transformation into the D=3 Z(2) gauge model on a simple
cubic lattice. This is the wanted result: the spectrum of the 
BCSOS interface should be the same as the spectrum of the Wilson
surface in the Z(2) gauge model.

Since we are interested in the full excitation spectrum of the Wilson
surface, we determined the low energy part of the full spectrum from
direct diagonalization of the transfer matrix of the BCSOS model.
A periodic boundary condition was used, which corresponds to the spectrum
of a periodic Polyakov line in the Z(2) gauge model. With a flux of period
L we used exact diagonalization for ${\rm L\leq 16}$, and the 
the Bethe Ansatz equations up to L=1024. 
The direct diagonalization was mainly used
for checks and establishing the pattern of level ordering.
The following picture emerges from the calculation for large L values
in the massless Kosterlitz-Thouless (KT) phase.
The ground state energy of the flux is given by

\begin{equation}
{\rm E_0(L)=\sigma_\infty\cdot L - \frac{\pi}{6L}c + o(1/L)~,}
\label{eq:bcsos1}
\end{equation}
where $\sigma_\infty$ is the string tension, c designates the conformal
charge, which is found to be c=1 to very  high accuracy, consistent
with the fact that we are in the KT phase.
The
o(1/L) term designates the corrections to the leading 1/L behavior;
they decay faster than 1/L.
At the critical point of the roughening transition, the corrections
can decay very slowly, like ${\rm 1/(lnL\cdot L)}$. Away from the
critical point, the corrections decay  faster than 1/L in power-like
fashion with some logarithmic corrections.

For each operator ${\rm O_\alpha}$ which creates states from the
vacuum with quantum numbers $\alpha$,
there is a tower excitation spectrum above the ground state,
\begin{equation}
{\rm E^\alpha_{j,j'}(L)=E_0(L) + \frac{2\pi}{L}(x_\alpha + j + j') + o(1/L)~,}
\label{eq:bcsos2}
\end{equation}
where the nonnegative integers j,j' label the conformal tower and
${\rm x_\alpha}$ is the anomalous dimension of the operator ${\rm O_\alpha}$.
The momentum of each excitation is given by
\begin{equation}
{\rm P^\alpha_{j,j'}(L)=\frac{2\pi}{L}(s_\alpha + j - j') ~,}
\label{eq:bcsos3}
\end{equation}
where ${\rm s_\alpha}$ is the spin of the operator ${\rm O_\alpha}$.

The surface excitation spectrum described by Eqs.~(\ref{eq:bcsos1},
\ref{eq:bcsos2}, \ref{eq:bcsos3}) is not a simple massless string
spectrum with obvious geometric interpretation. 
There are excitations with noninteger values of the anomalous
dimensions ${\rm x_\alpha}$ which continuously vary with the  
Ising coupling J. In fact,
we found an infinite sequence of operators which excite surface states
with fractional multiples of ${\rm 2\pi/L}$, instead of integer multiples
of ${\rm 2\pi/L}$, as expected in a naive string picture.
This sequence can be labelled by anomalous dimensions
\begin{equation}
{\rm x^G_{n,m} = \frac{n^2}{4\pi K} + \pi Km^2~,}
\label{eq:bcsos4}
\end{equation}
where n,m are nonnegative integers and the constant K depends in a known 
way on the BCSOS coupling constant J. The physical interpretation of
the rather peculiar excitations of the rough gapless surface will be discussed
elsewhere. Here it is sufficient to note that the spectrum is related
to a free compactified Gaussian field, but the field configuration 
allow for line defects, presumably related to dislocations of the
fluctuating rough surface.

\subsection{Surface Physics in the Three-Dimensional Z(2) Gauge Model}

First, we note that the SOS mapping of the Wilson
surface close to the roughening transition is not sensitive
to the group structure of the particular model in D=3 dimensions.
It is known that the SOS mapping in the 
Z(2) gauge model is accurate to a few percent
around the roughening transition. Assuming that the same is true in 
the SU(2) ${\rm QCD_3 }$ model, 
we should expect very
similar surface roughening behavior in the two models after
an appropriate rescaling of the gauge couplings  into the effective
coupling constant of the SOS model.

\begin {figure}[h]
\epsfxsize=4.8in
\epsfbox{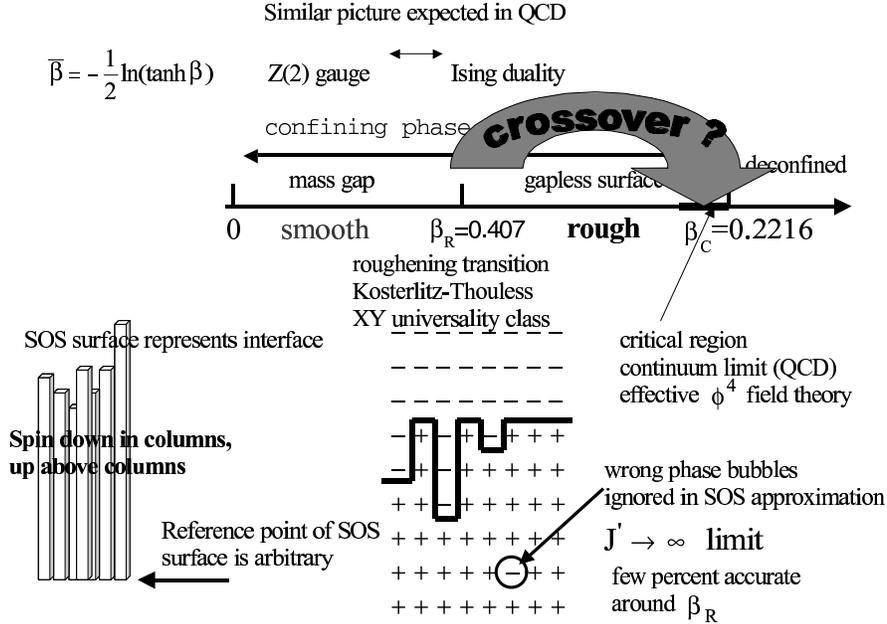}
\caption{The phase diagram of the bulk in the Z(2) gauge model
and the most important features of its Wilson surface are summarized
here. The horizontal axis represents ${\rm 1/\beta}$ on some arbitrary
scale. The zero temperature limit of the Ising representation corresponds
to ${\rm 1/\beta = 0}$ on the left end which is the strong coupling
limit of the Z(2) gauge model representation. Wrong phase bubbles are ignored
in the SOS representation of the surface which separates the spin-up region
from the spin-down region in Ising respresentation. These bubbles are
eliminated and the SOS approximation becomes exact 
in the ${\rm J'\rightarrow \infty}$ limit of the vertical Ising coupling.}
\label{fig:fig6}
\vskip -0.2in
\end{figure}

A great deal more is known about the Z(2)
gauge model in D=3 dimensions. 
Our goal is to draw analogy between the Z(2) model and the behavior of
${\rm QCD_3 }$ with SU(2) color which happens to exhibit 
most of the salient features of all the
other QCD simulations in the string formation region.
The phase diagram in the bulk and its
Wilson surface physics are summarized in Fig.~\ref{fig:fig6}. 
Based on the above remarks, we expect that
the analogy between SU(2) ${\rm QCD_3 }$ and Z(2)
remains useful {\em throughout the entire gapless rough phase}
of the Wilson surface, from the roughening 
transition to continuum string formation.

The key to the understanding
of the phase diagram is the well-known fact that the D=3 Z(2) gauge model
is dual to the D=3 Ising spin model. For the inverse gauge coupling
$\overline {\beta}$ of the Z(2) model the dual mapping  
onto ${\rm \beta = 1/kT}$ of the Ising model is given in Fig.~\ref{fig:fig6}.
In the bulk, the Z(2) model has a confining phase which corresponds to
the ordered phase of the Ising representation. This confined phase
ends at the bulk critical point ${\rm \beta_c=0.2216}$. In drawing
analogy with the SU(2) ${\rm QCD_3 }$ model, we should consider
the confined phase of the Z(2) model only. The ${\rm 1/\beta = 0}$
point on the left of the phase diagram 
maps into the  ${\rm g\rightarrow \infty}$ strong coupling
limit of ${\rm QCD_3 }$ whereas the ${\rm \beta_c=0.2216}$ Ising
critical point of the bulk maps into the ${\rm g\rightarrow 0}$
continuum limit of ${\rm QCD_3 }$.

\vskip 0.1in

\begin {figure}[h]
\vskip -0.2in
\epsfxsize=4.5in
\epsfbox{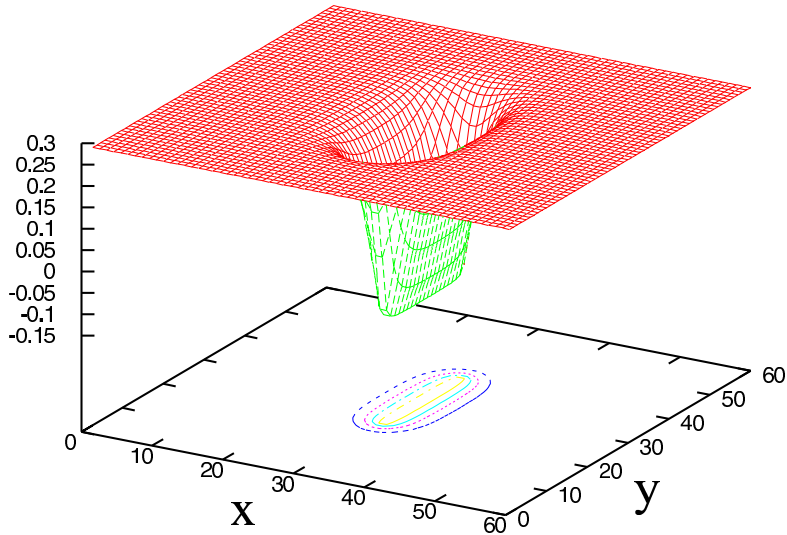}
\vskip -0.2in
\caption{The effective potential of the Schr\"odinger equation.}
\label{fig:fig7}
\vskip -0.1in
\end{figure}

\begin {figure}[h]
\vskip -0.2in
\epsfxsize=4.5in
\epsfbox{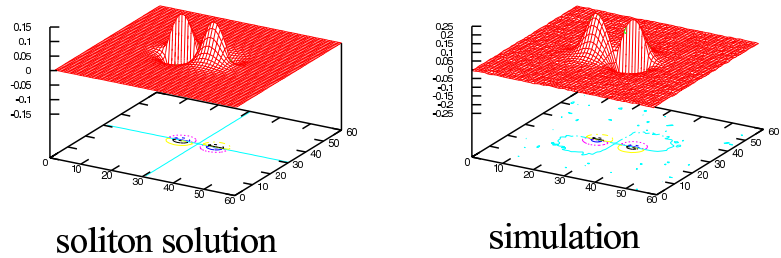}
\caption{The string soliton configuration corresponds to L=20
with clamped ends and bulk correlation length ${\rm \xi=1.87}$.}
\label{fig:fig8}
\vskip -0.2in
\end{figure}

\begin {figure}[h]
\vskip -0.25in
\epsfxsize=3.0in
\epsfbox{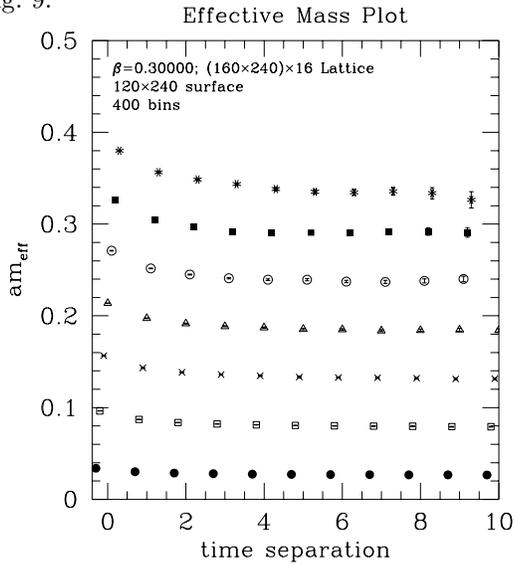}
\caption{The string spectrum of the Z(2) model is calculated here
for a location half way between the roughening transition and 
bulk criticality.}
\label{fig:fig9}
\vskip -0.35in
\end{figure}

\noindent{\bf The roughening transition region:}\newline
The confining flux sheet of the Wilson loop in the Z(2) gauge model
corresponds to the Ising interface in the dual representation.
The on-axis Z(2) Wilson surface exhibits a roughening
transition at ${\rm g_R}$ which maps into the inverse roughening
temperature  ${\rm \beta_R=0.407}$ of the Ising interface.
At the roughening point, the bulk is far from the critical coupling
${\rm g_c}$  of the Z(2) model which maps into the critical
coupling ${\rm \beta_c=0.2216}$ of the Ising representation.
This should correspond to the g=0 continuum limit in ${\rm QCD_3 }$.
The surface excitation spectrum in the roughening region
exhibits a surprisingly rich spectrum which is difficult
to interpret as an effective string theory based on a simple geometric picture
of the fluctuating interface. Some details of the spectrum were outlined
earlier in this section for the BCSOS representation 
where we presented the exact
solution. The two spectra should be essentially identical 
by universality arguments.
\vskip 0.05in
\noindent{\bf The continuum limit in the bulk:}\newline
When we move with the gauge coupling into the critical
region, depicted in Fig.~\ref{fig:fig6} as the close 
neighborhood of ${\rm \beta_c=0.2216}$ on the left, the SOS aproximation
is not valid anymore due to large fluctuations in the bulk.
A complementary description is 
expected to work in this region
in terms of a renormalization group improved
semiclassical expansion of the effective ${\rm \Phi^4}$
field theory, describing the critical region of the Z(2) model in Ising
representation.
The Wilson surface is described by a classical
soliton solution of the $\Phi^4$ field equations.
Excitations of the surface are given by
the spectrum of the fluctuation operator
\begin{equation}
{\mathcal M} = -\nabla^2 + {\rm U}^{\prime\prime}(\Phi_{\rm soliton})
\label{eq:soliton}
\end{equation}
where ${\rm U(\Phi)}$ is the field potential energy of the $\Phi^4$ field.
The spectrum of the fluctuation operator ${\mathcal M}$ of the
finite surface is determined
from a two-dimensional Schr\"odinger equation with a
potential of finite extent\cite{JK1}.
An example is shown in Fig.~\ref{fig:fig7} for L=30 separation
of the two fixed ends of the clamped Wilson surface at correlation
length ${\rm \xi=1.87}$ in the bulk.
In the limit of asymptotically large surfaces,
the equation becomes separable in the longitudinal and transverse
coordinates.
The transverse part of the spectrum is in close analogy with the
quantization of the one-dimensional classical $\Phi^4$ soliton.
There is always a discrete zero mode in the spectrum which is
enforced by translational invariance in the transverse direction.
The surface excitations have the form 
${\rm \Psi(x)\cdot exp(iqy),~q=\pi n/L,~n=1,2,3...}$, where the wavefuncion
${\rm \Psi(x)}$ represents the soliton zero mode. 
In Fig.~\ref{fig:fig8} we compare the second string state from the
semiclassical expansion with our direct simulation results.
In agreement with the semiclassical string soliton picture, we find 
a very accurate string spectrum even for correlation lengths
${\rm \xi\leq 1}$ as shown in Fig.~\ref{fig:fig9}. 

\begin {figure}[h]
\epsfxsize=4.0in
\vskip -0.5in
\epsfbox{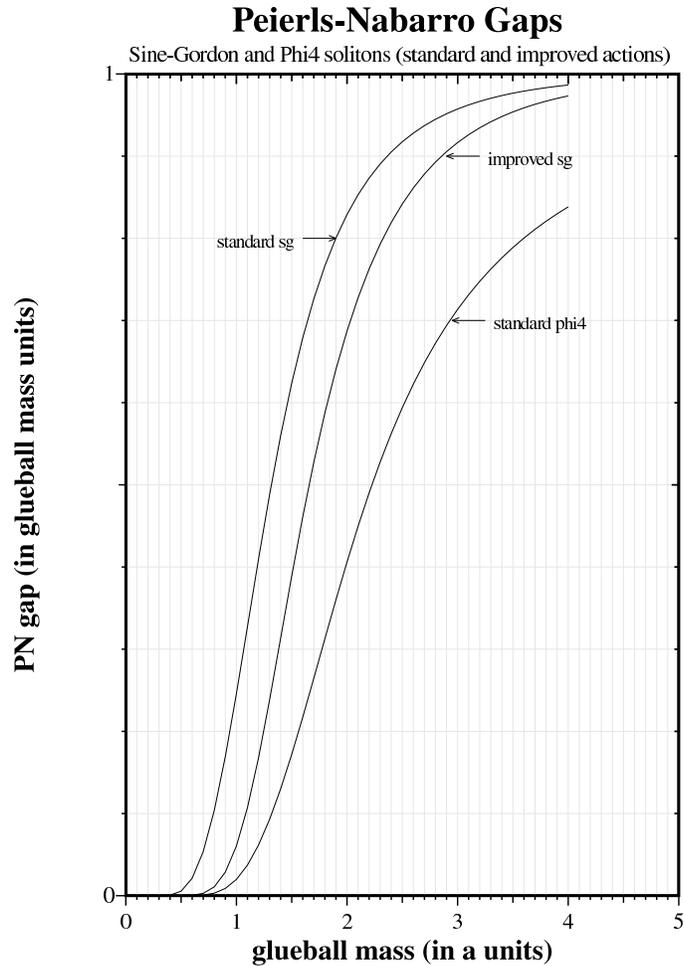}
\vskip -0.4in
\caption{Peierls-Nabarro gap in the string spectrum for various
scalar field theories in D=3 dimensions. The glueball mass is defined
as the inverse correlation length in the bulk.}
\label{fig:fig10}
\vskip -0.2in
\end{figure}

\newpage
The zero mode of
the soliton
is distorted on the lattice due to the explicit breaking of translational
invariance. In the leading semiclassical expansion this will generate
a gap which is the analogue of the Peierls-Nabarro gap in lattice
soliton physics. This gap is very significant for small correlation lengths,
as depicted in Fig.~\ref{fig:fig10}.
There is no trace of this gap in our simulations in Fig.~\ref{fig:fig9}.
The best guess is that tunneling and fluctuation effects, 
as the string soliton
moves in the periodic lattice potential, destroy the effect.

\section{What Is the Continuum Limit of QCD String Theory?}

Large Wilson loops in QCD represent the
euclidean time evolution of the gluon field generated by
the static quark-antiquark pair at large separation.
It is generally assumed that some confining flux develops
and the physical interpretation of its excitation spectrum
in lattice QCD requires a detailed  
understanding of critical phenomena on large Wilson
surfaces. 
In this section, we first outline our understanding
of the roughening transition of the Wilson surface in lattice QCD
and the crossover in surface criticality as we approach
the continuum limit in the bulk. This is mostly based
on our study of the Z(2) gauge model and the BCSOS model as presented
earlier.

\subsection{What is the String Interpretation of Roughening?}

If the $Q\overline Q$ pair is 
located along one of the principal axes on the lattice in some
spatial direction, 
the Wilson surface at strong coupling becomes {\em smooth} 
in technical terms. This implies the existence of a mass gap 
in its excitation spectrum, as seen for example in the
strong coupling tests of our simulation technology,
displayed in Fig.~\ref{fig:fig4}. 
The mass gap
is responsible for suppressing the fluctuations of the Wilson surface
away from its minimal area in the plane as determined by 
two principal lattice axes. 
One of the axes represents the space-like connection between
the static color sources and the other axis designates euclidean time.
From the viewpoint of statistical physics, the two directions are
equivalent, and we can talk about surface physics and its excitation
spectrum without further reference to the original physical
picture of the confined quark-antiquark pair and its gluon excitation
spectrum. 

As the coupling weakens, a roughening transition is expected in the surface at 
some finite gauge coupling ${\rm g=g_R}$ where the gap 
in the excitation spectrum 
vanishes. The correlation length in the surface diverges at the critical
point ${\rm g_R}$ of the roughening transition and it is expected to remain
infinite for any value of the gauge coupling when ${\rm g \leq g_R}$.
At the roughening transition, the bulk behavior remains far separated 
from the critical region of the continuum theory 
which is located in the vicinity of ${\rm g=0}$. Technically,
this implies that the surface will exhibit an infinite surface
correlation length
(rough surface) while the bulk correlation length is of the order one.
Based on strong
coupling series analysis and on the
approximate mapping of the Wilson surface fluctuations
into the solid on solid interface model,
surface roughening with the collapse of the mass gap at ${\rm g=g_R}$ is
expected to show the characteristics of the Kosterlitz-Thouless phase
transition. 

The low energy excitation spectrum of the Wilson surface 
for ${\rm g \leq g_R}$ and not far
from  ${\rm g_R}$, in the domain of the critical KT phase, should
be essentially identical to Eqs.~(\ref{eq:bcsos1},
\ref{eq:bcsos2}, \ref{eq:bcsos3}) of our BCSOS spectrum in ${\rm QCD_3}$.
This spectrum exhibits the features of a two-dimensional conformal
field theory with c=1 conformal charge. 
In ${\rm QCD_4}$ we do not expect qualitative changes.
The precise physical interpretation of the spectrum around the roughening
point in terms of a geometric string theory will require further work.
This is facilitated by the observation that the spectrum is equivalent
to that of the two-dimensional Gaussian scalar field on a circle,
including defect lines in the field configurations of the path
integral for the partition function.

\subsection{Crossover to continuum QCD}
Now, is the Kosterlitz-Thouless picture identical
to what we expect in continuum ${\rm QCD_3}$ string theory? 
As we have seen,
the Wilson surface in ${\rm QCD_3}$ is in the massless Kosterlitz-Thouless 
phase for gauge couplings weaker than the roughening coupling. 
Based on universality arguments, this alone
should determine the complete low-energy spectrum  of the surface.
However, as the coupling weakens below ${\rm g_R}$ and we take
the ${\rm g\rightarrow 0}$ continuum limit, an important
question arises. Do we expect a change in the structure of
the low energy
spectrum from the KT universality class into something else which should be
identified as the universality class of continuum QCD string theory?
This transition from the KT phase to continuum ${\rm QCD_3}$ string theory
should be particularly intriguing.
On one hand, the expected transition is quite  plausible, 
given the fact that we are sitting at ${\rm g_R}$ in the bulk 
which is far from
the critical region of the continuum limit. Why would this rough surface 
look identical to the continuum Wilson surface?
On the other hand, the Wilson
surface is unlikely to go back into a massive phase again as we
move towards ${\rm g=0}$. This would require a new critical point ${\rm g_c}$
somewhere between ${\rm g_R < g_c < 0}$ which is not likely.
The only plausible scenario is that the surface remains massless throughout
the ${\rm g_R \leq g \leq 0}$ region and its critical behavior will cross over
from the Kosterlitz-Thouless class into the universality class
of continuum QCD string theory whose precise description remains
the subject of our future investigations.

\section*{Acknowledgments}
One of us (J.~K.) would like to acknowledge valuable discussions
with P.~Hasenfratz, K.~Intriligator, F.~Niedermayer, 
J.~Polchinski, S.~Renn, U.-J.~Wiese,
and J.-B.~Zuber. J.~K. is also thankful
to the organizers of the workshop who created a stimulating
atmosphere throughout the meeting.
This work was supported by the U.S.~DOE, Grant No.\ DE-FG03-97ER40546.

\section*{Appendix}

Three exact quantum numbers which are based on the symmetries of the problem
determine the classification scheme of the gluon excitation spectrum.
We adopt the standard notation from the physics of diatomic molecules
and use $\Lambda$ to denote the magnitude of the eigenvalue of the projection
$\vec{J_g}\!\cdot\hat{\bf r}$ of the total angular momentum $\vec{J_g}$
of the gluon field onto the molecular axis $\hat{\bf r}$. The capital Greek
letters $\Sigma, \Pi, \Delta, \Phi, \dots$ are used to indicate states
with $\Lambda=0,1,2,3,\dots$, respectively.  The combined operations of
charge conjugation and spatial inversion about the midpoint between the
quark and the antiquark is also a symmetry and its eigenvalue is denoted by
$\eta_{CP}$.  States with $\eta_{CP}=1 (-1)$ are denoted
by the subscripts $g$ ($u$).  There is an additional label for the
$\Sigma$ states; $\Sigma$ states which
are even (odd) under a reflection in a plane containing the molecular
axis are denoted by a superscript $+$ $(-)$.  Hence, the low-lying
levels are labelled $\Sigma_g^+$, $\Sigma_g^-$, $\Sigma_u^+$, $\Sigma_u^-$,
$\Pi_g$, $\Pi_u$, $\Delta_g$, $\Delta_u$, and so on.  For convenience,
we use $\Gamma$ to denote these labels in general.

The gluon excitation energies $E_\Gamma(\vec{r})$ were extracted from 
Monte Carlo estimates of generalized large Wilson loops.
Recall that the well-known
static potential $E_{\Sigma_g^+}(r)$ can be obtained from the large-$t$
behaviour $\exp[-tE_{\Sigma_g^+}(r)]$ of the Wilson loop for a rectangle
of spatial length $r$ and temporal extent $t$.	In order to determine the
lowest energy in the $\Gamma$ sector, each of the two spatial segments of the
$r\times t$ rectangular Wilson loop must be replaced by a {\em sum} of
spatial paths, all sharing the same starting and terminating sites,
which transforms as $\Gamma$ under all symmetry operations.  The easiest
way to do this is to start with a single path ${\cal P}_\alpha$, such as
a staple, and apply the $\Gamma$ projection operator which is a weighted
sum over all symmetry operations; this yields a single gluon operator in
the $\Gamma$ channel.  Different gluon operators correspond to different
starting paths ${\cal P}_\alpha$.  Using several (in some channels as 
many as 40)
different such operators then produces a matrix of Wilson loop correlators
$W_\Gamma^{ij}(r,t)$.

Monte Carlo estimates of the $W_\Gamma^{ij}(r,t)$ matrices were obtained
in several simulations performed on our Beowolf class UP2000 Alpha cluster
using an improved gauge-field
action\cite{peardon}.  The couplings $\beta$, input aspect ratios $\xi$,
and lattice sizes for each simulation will be listed in
tables of our forthcoming publication\cite{JKM4}.  
Our use of anisotropic lattices in which
the temporal lattice spacing $a_t$ was much smaller than the spatial
spacing $a_s$ was crucial for resolving the gluon
excitation spectrum, particularly
for large $r$.	The couplings in the action depend not only on the QCD
coupling $\beta$, but also on two other parameters: the mean temporal
link $u_t$ and the mean spatial link $u_s$.  Following earlier
work\cite{peardon},
we set $u_t=1$ and obtain $u_s$ from the spatial plaquette.  We correct
$a_s/a_t=\xi$, the input or bare anisotropy, in all of our calculations,
by determining the small radiative corrections to the anisotropy as finite
lattice spacing corrections which vanish in the continuum limit.

To hasten the onset of asymptotic behaviour, iteratively-smeared spatial
links\cite{peardon} were used in the generalized Wilson loops.
A single-link procedure was used in which each spatial link variable
$U_j(x)$ on the lattice is mapped into itself plus a sum of its four
neighbouring (spatial) staples multiplied by a weighting factor $\zeta$.
The resulting matrix is then projected back into SU(3).	 This mapping
is then applied recursively $n_{\zeta}$ times, forming new smeared links
out of the previously-obtained smeared links.  Separate measurements
were taken for each smearing; cross correlations were not determined.
The temporal segments in the Wilson loops were constructed from
thermally-averaged links, whenever possible, to reduce statistical noise.

\end{document}